\documentclass[apj]{emulateapj}

\usepackage{graphicx}      
\usepackage{natbib}

\bibpunct{(}{)}{;}{a}{}{,}
\begin{document}

\title{${\rm C}^{18}$O Depletion in Starless Cores in Taurus}
\author{
Amanda Brady Ford\altaffilmark{1},
Yancy L. Shirley\altaffilmark{2,3}}

\altaffiltext{1}{Steward Observatory, University of Arizona,
                933 N. Cherry Ave., Tucson, AZ 85121; aford@as.arizona.edu}
\altaffiltext{2}{Steward Observatory, University of Arizona,
                933 N. Cherry Ave., Tucson, AZ 85121; yshirley@as.arizona.edu}
\altaffiltext{3}{Adjunct Astronomer at the National Radio Astronomy Observatory}

\begin{abstract}
We present here findings for ${\rm C}^{18}$O depletion in eight starless cores in Taurus: TMC-2, L1498, L1512, L1489, L1517B, L1521E, L1495A-S, and L1544. We compare observations of the ${\rm C}^{18}$O J=2-1 transition taken with the ALMA prototype receiver on the Heinrich Hertz Submillimeter Telescope to results of radiative transfer modeling using RATRAN. We use temperature and density profiles calculated from dust continuum radiative transfer models to model the ${\rm C}^{18}$O emission. We present modeling of three cores, TMC-2, L1489, and L1495A-S, which have not been modeled before and compare our results for the five cores with published models.  We find that all of the cores but one, L1521E, are substantially depleted. We also find that varying the temperature profiles of these model cores has a discernable effect, and varying the central density has an even larger effect. We find no trends with depletion radius or depletion fraction with the density or temperature of these cores, suggesting that the physical structure alone is insufficient to fully constrain evolutionary state. We are able to place tighter constraints on the radius at which ${\rm C}^{18}$O is depleted than the absolute fraction of depletion. As the timeline of chemical depletion depends sensitively on the fraction of depletion, this difficulty in constraining depletion fraction makes comparison with other timescales, such as the free-fall timescale, very difficult.
\end{abstract}

\begin{subjectheadings} 
sstars: formation -- ISM: abundances -- ISM: clouds -- ISM: molecules -- Individual Objects: TMC2, L1498, L1512, L1489, L1517B, L1521E, L1495AS, L1544
\end{subjectheadings}

\maketitle

\section{Introduction}

Understanding the transformation of a starless core into a protostellar object is the first step in understanding the overall process of low-mass star formation. A key component in understanding low-mass star formation is determining the evolutionary states of starless cores. The competing models for star formation - ambipolar diffusion and gravo-turbulent fragmentation - each have very different timescales \citep[see, e.g.,][]{Shuetal1987,gravfrag, maclow, Evansreview, hartmannetal}. 

Ambipolar diffusion posits that collapse of a starless core is slowed by neutrals slipping relative to ions and the decay of magnetic fields \citep{Shuetal1987}. The ambipolar diffusion timescale is approximately ${5}\times{10}^{6}$ years, assuming an ionization fraction of ${10}^{-7}$. Gravo-turbulent fragmentation proposes that large-scale turbulence causes fragmentation of individual cores, which then collapse on a free-fall timescale. For ${\rm H}_{2}$ densities of ${5}\times{10}^{5}$ ${\rm cm}^{3}$, that implies a timescale of $5\times{10}^{4}$ years. 

One approach is to use chemical evolution, or the process by which the chemistry of a core changes over time, as another piece of information to distinguish between these theories.  In the cold ($\leq$ 10K), well-shielded interiors of dense starless cores, significant freeze-out (depletion) of molecular species such as CO, CS, CCS, and HCO$^+$ are observed (e.g., \cite{Tafalla416}).  The more evolved cores are expected to have more freeze-out, or depletion \citep{Tafalla416}. For example, \cite{Crapsi2007} found that in L1544, 93\% of the ${\rm C}^{17}$O has frozen out in the center of this very centrally condensed core (central density of $8 \times 10^5$ cm$^{-3}$). Another example is the work of \cite{Redmanetal} toward L1689B (central density of ${2.8}\times{10}^{5}$ cm$^{-3}$ \citep{Bacmannetal}) where 90\% of CO is depleted after ${4}\times{10}^{4}$ years, or half the free-fall timescale at that density. However, this is a lower limit, as they assume a sticking coefficient of unity, and a constant density over time (which, as they point out, is unreasonable - the density would increase over time). 

The state-of-the-art theoretical studies are chemodynamical models that use a chemical network at each time step during the collapse of a starless cores to calculate abundances.  For instance, the models of \cite{Lee2004} shows significant chemical depletion happening at roughly ${10}^{6}$ years. It is therefore theoretically possible that the depletion profile can be used to distinguish between ambipolar diffusion and gravo-turbulent fragmentation.

In this paper, we map the J=2-1 transition of ${\rm C}^{18}$O toward a sample of eight starless cores in the nearby Taurus molecular cloud. By comparing observed integrated intensities with intensities obtained from radiative transfer modeling, we place constraints on the amount of depletion in terms of how much CO is depleted (fraction of depletion ${f}_{d}$), and where this occurs (the depletion radius ${r}_{d}$ ).  Our work is unique because we use a larger sample set (8 cores), all located in the same molecular cloud and are then able to compare the amount of depletion in these cores to evolutionary parameters, such as central density. We do not assume a constant ${\rm T}_{g}$=10~K profile for our models, as has been done in previous work, and investigate degeneracies in choice of temperature versus choice in abundance. We also investigate possible effects of the gas temperature and the dust temperature not being equal. Unless otherwise specified by ${\rm T}_{dust}$, temperatures referenced are the kinetic gas temperature,${\rm T}_{g}$ .

This paper is organized as follows. In \S~2 we discuss our sample selection and motivation. We also describe our observations and modeling methodology, including the implementation of the RATRAN (\cite{ratran}) radiative transfer code. In \S~3 we test our modeling procedure, including potential degeneracies of various parameters, and present results for each individual core. In \S~4 we present our analysis and in \S~5 we give our conclusions.

\section{Motivation and Procedures}

\subsection{Sample Selection and Motivation}

Our sample consists of cores in Taurus, at approximately 140 pc  \citep{VLA}, so they are easy to compare at the same spatial scale. We chose this particular set because they span over two orders of magnitude in central density (see Table 1). Figure \ref{dustemiss} shows observations of the dust continuum emission at 850~\micron\  taken with SCUBA, as well as ${\rm C}^{18}$O 2-1 integrated intensity maps. These cores have at least some rough central dust continuum concentration. 

Given a constant temperature profile (see \S~3.2 for how this changes if temperature varies), and a constant [${\rm C}^{18}$O]/[${\rm H}_{2}$], one would expect to observe more intense ${\rm C}^{18}$O  emission in the center that at outer radii. For many of these cores, that is not what is observed, as seen in Figure \ref{dustemiss}. Given the high central densities of these cores, the ${\rm C}^{18}$O could be depleted by freeze-out onto dust grains. The greater the density, the more freeze-out one would expect, so this freeze-out would occur preferentially in the centers of the cores  \citep{Goldsmith}. We shall model this effect with radiative transfer.

\subsection{Observations}
We took observations of the  J=2-1 transition of ${\rm C}^{18}$O (219.56036 GHz) during February 5 - 8, 2007, using the sideband-separating ALMA 1mm (Band 6) prototype receiver on the 10-m Heinrich Hertz Submillimeter Telescope (HHT).  The beam size, (full-width half max; FWHM) at the observed frequencies is 34$^{\prime\prime}$. The main beam has been measured to be Gaussian to detectable levels using maps of Mars (J. Bieging, private communication, 2008). For the ALMA prototype receiver, the sidelobes are insignificant, as those are down by at least 20dB. The C$^{18}$O line was tuned to the lower sideband with a 6.0 GHz IF.  Receiver noise temperatures were between $90 - 100$ K with system temperatures $< 250$ K.  Typical sideband rejections were greater than $13$ dB and were ignored in the subsequent calibration.  The ALMA 1mm prototype receiver is both sideband separating and dual polarization resulting in 4 intermediate frequency outputs (Vpol LSB, Hpol LSB, Vpol USB, and Hpol USB) to the spectrometers.  Unfortunately, we were only able to use one polarization (Vpol) of the lower sideband since there was only one high spectral resolution backend ($\Delta \nu < 100$ kHz) available in February 2007.  A Chirp Transform Spectrometer with $32$ kHz spectral resolution was used for all C$^{18}$O observations.  The HHT uses the standard Chopper Wheel Calibration Method \citep{pandb}.  The data were baselined and calibrated using the CLASS reduction package.  Mapping observations were obtained on a 10$^{\prime\prime}$ grid covering $2^{\prime} \times 2^{\prime}$.  Each position was observed for 2 minutes of integration time in position-switching mode.  OFF positions were checked for C$^{18}$O emission.  Calibration scans of Saturn were made every two hours after pointing and focusing the telescope.  The final reduced spectra were placed on the $T_{mb} = T_A^* / \eta_{mb}$ scale with an average main beam efficiency of $\eta_{mb} = 0.74$. 

Rms errors for the observed integrated intensity points generally ranged from 0.1 to 0.2, the rms error of the observed position with the largest $\sigma_{I}$ for each core, is given in Table 1. Errors are defined as:

\begin{equation}
\sigma_{I}  = \sqrt {\Delta V_{line} \Delta V_{channel}} * \sigma_T
\end{equation}

where  $\Delta V_{line}$  is the velocity interval over the whole line, from line wing to line wing. $\Delta V_{channel}$ is the channel width, of 0.157 km/sec, and $\sigma_T$ is the baseline rms error, assuming a linear baseline.

\subsection{Modeling}

A radiative transfer calculation is appropriate for this work, as computing a column density from observations requires the assumption of a single excitation temperature throughout the core, which is incorrect. Also, because we are in non-LTE, the kinetic temperature is not equal to the excitation temperature along the entire line of sight.  Therefore, we use the 1-D version of the Monte Carlo radiative transfer code, RATRAN, developed by \cite{ratran}, to generate radial profiles of ${\rm C}^{18}$O emission.  RATRAN uses Accelerated Lambda Iteration to efficiently reach convergence between the level populations and calculations of $J_\nu$, the mean intensity of the radiation field. 

RATRAN inputs for our models include:

\begin{enumerate}
\item ${\rm C}^{18}$O molecular collision rate file from the LAMBDA database (see \cite{moldat}).

\item Dust grain properties: We use OH5, see \cite{Ossenkopf}, Table 1, column 5. These opacities typically give the best fit in starless core dust continuum radiative transfer models (see Shirley et al. 2005).

\item Radial points: We used 100 radial shells, from 20~AU to 30,000~AU, with equal logarithmic spacing.

\item ${\rm H}_{2}$ density profiles: The density profiles are constrained by the Shirley \& O'Malia dust continuum radiative transfer models (see also \cite{Shirleyetal2005} for methodology).  The models assume isothermal, static Bonnor-Ebert spheres with various central densities \citep {Bonnor,Ebert} which are derived under assumptions of pressure-bounded hydrostatic equilibrium.  See Figure \ref{alldensity} for our density profiles.  While the B-E spheres are isothermal, their resulting temperature profile from the dust continuum radiative transfer calculation is non-isothermal; however, \cite{Evansetal2001} found that the correction to n(r) for the non-isothermal case is negligible for low-mass starless cores. See \cite{Shirleyetal2005} for further discussion.

Our central densities are not identical to those used in other published work, specifically in \cite{Kaisa}, because Shirley \& O'Malia use a finer central density grid in the modeling. We also assume a ${\rm H}_{2}$ ortho-para ratio of 3:1, as in \cite{Flowers}, and find that our results do not depend strongly on o:p ratio.

Running these density profiles through RATRAN requires choosing a cutoff radius, beyond which we assume ${\rm H}_{2}$ (and, by extension ${\rm C}^{18}$O) has its standard ISM value (see below). The cutoff radius is necessary for modeling; it is not physically motivated. In reality, the edge of a core is difficult to define, and there is more likely to be a gradual decline in density versus a sharp discontinuity between the core and surrounding ISM, we investigated how sensitive our results were to this choice. This was especially important given that work by \cite{Onishi1996} found large amounts of ${\rm C}^{18}$O in the ambient Taurus molecular cloud, not necessarily associated with starless cores. We chose 30,000~AU as a reasonable outer radius for our models, consistent with the 0.1~pc typical core size given by \cite{Tafallareview}.  For the ${\rm C}^{18}$O observations, we mapped only to $\pm$60$^{\prime\prime}$ (corresponding to 8400~AU at 140~pc) from the dust continuum peak, as the dust maps were restricted by chopping ($\theta_{chop}/2 = 60^{\prime\prime}$). Figure \ref{nofr} shows that most of the contribution to the total column density occurs in the center parts of the cloud, suggesting our choice of outer radius is reasonable.

\item ${\rm C}^{18}$O abundance profiles: We assume a step function, wherein ${\rm C}^{18}$O is depleted by some fraction ${f}_{d}$ at some radius ${r}_{d}$. We define:

\begin{equation}
{f}_{d}\equiv\frac{([{\rm C}^{18}O]/[{\rm H}_{2}])_{ISM}}{([{\rm C}^{18}O]/[{\rm H}_{2}])_{interior~to~{r}_{d}}}
\end{equation}
We vary ${f}_{d}$ from 1.0 (no depletion) to $10^3$ in steps of ${2}^{n}$, with n varying from 0 to 10. We vary ${r}_{d}$ from 1,250~AU to 30,000~AU in steps of 1,250~AU.

Exterior to ${r}_{d}$, [${\rm C}^{18}$O]/[${\rm H}_{2}$] has its standard ISM abundance, which we define in this work as  $1.6\times 10^{-7}$, herafter referred to as the "canonical abundance". This is slightly lower than work by \cite{Frerkingetal}, their $1.7\times 10^{-7}$ vs our $1.6\times 10^{-7}$;  we adjust it to be consistent with work by \cite{WilsonRood}, who find a slightly lower $^{18}$O/$^{17}$O ratio. Using $\chi^2_r$ fitting, we find an ${r}_{d}$ and ${f}_{d}$ that best match both the observed spectra for the central region and the observed integrated intensity of ${\rm C}^{18}$O emission across the core.
Previous work on two of our cores, L1498 and L1517B, by \cite{Tafalla416} has shown that the abundance drop is quite sharp and therefore that a step function is reasonable (see also \cite{Leeetal}). See Figure \ref{covsr} for our ${\rm C}^{18}$O density profiles.

\item Temperature profile of ${\rm C}^{18}$O:  The temperature profiles are self-consistently calculated from dust continuum radiative transfer models of the cores.  The dust continuum emission from the eight cores in this study were modeled as part of a larger sample of starelss cores that were observed with SCUBA (Shirley \& O'Malia, in prep).  A one dimensional dust continuum radiative transfer code, CSDUST3, \cite{Eganetal1988} is used to match submillimeter intensity profiles and spectral energy distributions (SEDs).  The heating in the dust continuum models is supplied by the Interstellar Radiation Field (ISRF) which is allowed to vary in overall strength and the degree of extinction (see Shirley \& O'Malia, in prep.)  For our purposes in RATRAN, we approximate the T(r) profile as the same for the gas kinetic temperature and the dust temperature. Work by \cite{Kaisa} has indicated that this assumption holds true for densities above $10^4$ in L1498, above $10^3$ in L1512, and above $5\times 10^4$ in L1544. Discussion of the implications of these assumptions can be found in \S3.2.  Figure \ref{alltemp} shows the temperature profile used in each core.  

\item Velocity profile: We use a static velocity field with turbulent and thermal line broadening described by the Doppler b parameter = 0.2, since that quantity fits the observed line widths. Late stages of collapse would produce line widths broader than those observed. Work by \cite{Meyers} has indicated that the shape of the B-E sphere density profiles do not change substantially even in the collapsing case, until late in the collapse history.  

\end{enumerate}

	After the model has been run through RATRAN, we use the Miriad image analysis package \citep{Miriad} to convolve the outputs with our 34" beam size, and integrate over velocity to determine the integrated intensity in K km/s.

Before we began any of our modeling, we first attempt to replicate previously published results, as a check on our procedures for running the radiative transfer code. We were graciously provided density and temperature profiles by Mario Tafalla from work on L1521E \citep{TafallaL1521E}.   See Fig \ref{Tafallaprofilecompare} for those profiles, as compared to our inputs for the same core. We were able to replicate their published best fit curves through our analysis, even though they used a different radiative transfer code \citep{Bernes}. This indicates that any differences in our results are based on differences in real physical inputs, such as density and temperature, and not differences in modeling procedures. L1521E results are further discussed in \S~3.3.

\section{Discussion}

We begin our modeling by using model temperature and density profiles as discussed above, and then varying ${r}_{d}$ and ${f}_{d}$, as in previous work to find best fits. ${r}_{d}$ and ${f}_{d}$ are considered free parameters and are fully explored as such. As these parameters represent the amount of depletion in each core, which are then used to make chemical evolution time scale arguments, it is important to understand how sensitive the results are to the assumptions that go into their calculation. We explore how much these calculated values of ${r}_{d}$ and ${f}_{d}$ would change if the RATRAN inputs changed. Of the seven RATRAN inputs described in \S~2.3, we explore factors influencing the canonical abundance and temperature profiles of ${\rm C}^{18}$O.  Inputs for molecular collision rates, dust grain properties, and velocity profiles are already as well constrained as they are likely to be for now, and there is good agreement in the literature, so we do not explore those further. ${\rm H}_{2}$ density profiles are well-constrained by dust continuum modeling, and choice of cutoff radius is of little consequence to the model column densities.  

We choose to explore the canonical abundance of ${\rm C}^{18}$O because there is a large amount of disagreement in the literature over this value, and as we ourselves have no good way of independently verifying its correct value. We choose also to explore the sensitivity to the temperature profile, because while we self-consistently calculate a varying profile, other work assumes a constant profile. Furthermore, there are physically motivated reasons to believe that choosing different values of canonical abundance and/or temperature than described in the previous section would yield different best-fit ${r}_{d}$ and ${f}_{d}$. The point of this analysis is not to find an absolute best fit for this multi-degree parameter space, since input data is too limited to make such a finding meaningful. The point is to see how much caution is needed - if we use ${r}_{d}$ and ${f}_{d}$ as proxies for chemical evolution, and if these proxies are extremely sensitive to our input assumptions, then our work and future work needs to be mindful of this.

\subsection{Variations in the ISM ${\rm C}^{18}$O abundance}

As discussed in \S~2.3, we assume that, in the absence of freeze-out, the ratio of [$\rm{C}^{18}$O]/[${\rm H}_{2}$] is fixed for all of our cores at the canonical abundance of $1.6\times10^{-7}$. However, not everyone agrees that this is the correct value to use. A comparison of published work in the field shows that values of the canonical [$\rm{C}^{18}$O]/[${\rm H}_{2}$] abundance differs by a factor of seven, from $0.7\times10^{-7}$ \citep {TafallaL1521E} to $1.5\times10^{-7}$ \citep{Tafalla416} to $4.8\times10^{-7}$ \citep{Leeetal}. It may be that more than one of these values is correct, as the canonical abundance could even vary across the Taurus region. This may be a reasonable possibility in Taurus, given the large scale structure of filaments, cavities, and rings and subsequent variations in the radiation field found by \cite{GoldsmithLSS}. 

Because of the wide range of published values for this particular input parameter, we choose to explore it in further detail by varying the canonical abundance and seeing how that affected ${f}_{d}$ and ${r}_{d}$. A full exploration of all possible values of canonical abundance is beyond the scope of this work, but we felt it necessary to at least gain a quick understanding of the nature of the impact. We ran models with all of the same input parameters described in \S~2.3, but for half the canonical abundance, as well as twice the canonical abundance. In changing the canonical abundance, some of the cores were best fit by entirely different values of ${f}_{d}$ and ${r}_{d}$ than for an abundance ratio of $1.6\times10^{-7}$ (see next section for further discussion). To our knowledge even this cursory exploration of variances in canonical abundance has not been done before.  Further constraints on canonical abundance are needed to be confident that  ${f}_{d}$ and ${r}_{d}$ values generating lowest $\chi^2_r$ for choice of one canonical abundance have physical meaning. More details can be found in \S~3.3.

\subsection{Sensitivity to Temperature Profiles} 

We also explored how sensitive best fit ${f}_{d}$ and ${r}_{d}$ values were to choice of temperature values. We started this investigation by using all of the same input parameters as before (see \S~2), but changing the temperature profiles. We explored both temperature profiles that vary and those remaining constant across the core. Our objective was not, in this way, to constrain the temperature profiles of these cores but to understand how sensitive results are to this parameter. We believe we are able to constrain well the temperature profiles from dust continuum radiative transfer modeling, described in \S~2.3. Additionally, a varying temperature profile, as used in our work, makes physical sense: the areas near the center of the core, being shielded from the interstellar radiation field, should be colder than those on the outside (\cite{Zucconi}). Dust continuum modeling of our sample of cores suggests that while the temperature variation is only a few Kelvin, this means that the outer radius of the core is up to 30\% warmer than the center (Figure \ref{alltemp}). 

However, much of the other work in this field has assumed that the temperature remains constant throughout the core, with a typical value of T=10~K for all radii.  Given that our work does not make this assumption, we explore how temperature profiles could affect which ${f}_{d}$ and ${r}_{d}$ best fit the observational data.  It is, of course, the combination of temperature and density that produce the observed intensity. If a core is cooler than assumed, then one would require a higher density to match the observations, and vice versa. This would lead one to believe a core is less depleted than if one started with a higher temperature. 

We ran a representative set of models for B-E spheres with central densities of $3\times 10^5$ and $2.5\times 10^4$ (see Figures \ref{nc3e5plus1} and \ref{2pt5e4}). These Figures show the ${\rm C}^{18}$O 2-1 intensity versus radius for various combinations of {\it ${r}_{d}$} and {\it ${f}_{d}$}, using various temperature profiles. We plotted three constant temperature profiles: T=7~K, T=10~K, and T=13~K, as well as a temperature profile that varied with radius between 8 and 13~K. For Fig \ref{nc3e5plus1}, note how close the red line (temp=10~K) is to the black line (the varying temperature profile). Since the varying temperature profile goes from 8-12~K, it seems one can simply take the average of those numbers and use that as the single temperature. Indeed, the shapes of all the curves are fairly similar.  This is consistent with work by \cite{Tafalla416}, who found that modeling with isothermal profiles can well reproduce the data. However, the exact choice of that isothermal temperature is important. From Figures \ref{nc3e5plus1} and \ref{2pt5e4}, one can see that not all cores can be modeled at T=10~K. One would need further information, such as self-consistent temperature modeling from dust emission, as in this work, to know which isothermal temperature to use. Figure \ref{alltemp} shows the temperature profiles we adopt for each core in the rest of this work.

Another common assumption is that the gas and dust temperatures are well coupled (we make that assumption in our work here). At high densities this is true, as collisions between the gas and dust temperatures result in equal kinetic gas and dust temperatures. At low densities, however, as in those found on the outside of the core, this is not necessarily the case. Indeed, recent work by \cite{Kaisa} has shown that gas and dust temperatures can vary substantially in those conditions. However, precisely because $T_d \neq T_{gas}$ only on the outside of the core, in the lowest density areas that contribute least to the column density (see earlier section), we do not expect this to be a large effect. We were graciously provided gas and dust temperature profiles by Young et al for three cores for which they did detailed energetics calculations: L1498, L1512, and 1544. We used our same model density profiles but their calculated gas and dust temperature. See \S~3.3 for further discussion.

We will see in section \S~3.3 that in many cases we find different ${f}_{d}$ and ${r}_{d}$ values for many of the cores than previously published. Based on the above analysis, we believe some of this discrepancy is likely due to our temperature profiles being different from those used in other work. As another quick check, we simultaneously varied both the canonical abundance and the temperature for a small slice of parameter space. We found that there can also be a degeneracy in the abundance and the temperature used, as in Fig \ref{L1521Edegeneracies}, which worsens the problem. One can well fit the data by assuming a higher canonical abundance and a lower temperature, or a lower canonical abundance with a higher temperature. 

\subsection{Summary of Findings for Each Core} 

Figure \ref{azavcan} plots the C$^{18}$O integrated intensity $\chi^2_r$ for depletion profiles with combinations of ${f}_{d}$ and ${r}_{d}$.  Due to the high scatter in the observed intensity points, we azimuthally averaged our observed points before comparing them to our 1-D model.  We divided the data into 10" annuli around the center core, and averaged the observed and model points over that range. We calculated a $\chi^2_r$ as follows: 

\begin{equation}
\chi^{2}_r\equiv\sum\frac { ({\rm I}_{obs}- {\rm I}_{mod})^{2}}{\sigma_{obs}*({\rm number \,of \,annuli - 1}) }
\end{equation}

We define ${\rm I}_{obs}$ as the mean intensity of each of the observed points, for each 10" annuli. We define ${\rm I}_{mod}$ as the intensity of the middle model point in the same annuli. We define  $\sigma_{obs}$ as the variance in the intensity of the observed points for each annuli. For annuli with only one observed points (for some of the cores, there was only one observed data point in the 0-10" annuli), we define the variance as simply the error on that one point. Errors for these individual points are defined in equation 1. We have six annuli, going from 0 to 60". 

For TMC-2, L1521E, L1495A-S, and L1544, no combination of ${f}_{d}$ and ${r}_{d}$ with a canonical abundance gave a $\chi^2_r<$1. However, as seen in Figure \ref{azavtwicecan}, increasing the canonical abundance by a factor of two was able to give $\chi^2_r<$1 values under 1, except for L1544 (discussed below). L1512 and L1517B had combinations of ${f}_{d}$ and ${r}_{d}$ that gave $\chi^2_r<$1 for both the canonical abundance and twice the canonical abundance, however using twice the canonical abundance ultimately gave lower $\chi^2_r$'s. None of the cores had lower $\chi^2_r$ when the canonical abundance was halved, so we do not favor this possibility. In addition to these $\chi^2_r$ maps, we also calculate $\chi^2_r$ of the line profiles for the central 10$^{\prime\prime}$ regions of these cores, as shown in Figure \ref{centerchisqcontour}. Figure \ref{centerchisq} shows observations of the central regions with best fit models (see Table 2).  Both $\chi^2_r$ calculations have significant overlap between the best fit values of ${f}_{d}$ and ${r}_{d}$.  While all of these models use the temperature profile calculated from dust continuum radiative transfer models, Figure \ref{Kaisa} plots the integrated intensity $\chi^2_r$ for three cores using the temperature profiles from the energetics calculations of \cite{Kaisa}, where $T_d \neq T_{gas}$ for all radii. This gives very similar results to Figure \ref{azavcan},  consistent with the conclusion that using uncoupled gas and dust temperatures does not produce a substantial effect on the C$^{18}$O J=$2-1$ emission. 

Inspection of Figures \ref{azavcan} and \ref{azavtwicecan} shows that there is some degeneracy between exactly where the depletion occurs and by exactly how much. In many cases it is easier to constrain ${r}_{d}$ than ${f}_{d}$. These figures also show the difficulty in constraining ${f}_{d}$, as a range of  ${f}_{d}$ values all give $\chi^2_r$ $<$1.  Fig \ref{ivsrfits} shows the observations for each core, as well as models with various $\chi^2_r$ values. Despite this degeneracy, it is clear that L1521E has low depletion while cores such as L1512 and
L1517B, for example, have higher depletion. The best fit ${f}_{d}$ and ${r}_{d}$ values for each source are listed in Table 2. Details for each core are given below.

\begin{enumerate}

\item TMC2: this is a difficult core to model, since, based on dust emission (see Figure \ref{dustemiss}), it is not as centrally condensed, and is non-azimuthally symmetric. In the case of canonical abundance, it is best matched by a model profile with no depletion. However, even this profile has a fairly high $\chi^2_r$. In order to get the lowest $\chi^2_r$, we had to double the ${\rm C}^{18}$O ISM abundance. The lowest $\chi^2_r$ for TMC-2 is a depleted model with twice the canonical abundance.

\item L1498:  We find this core is well-modeled with canonical abundance of  ${\rm C}^{18}$O with significant depletion. We find similar depletion to work by \cite{Willacyetal} , which sets a lower limit on the depletion fraction at 8. Work by \cite{Tafalla416} found ${\rm C}^{18}$O  to be entirely depleted (${f}_{d}$$>$ 1000) inside a radius of 9,940~AU. We find a ${r}_{d}$ of 7,500~AU but only a factor of 16 depletion. However, as shown in Fig \ref{azavcan}, we find that our findings and those of \cite{Tafalla416} to have very similar $\chi^2_r$. The slight discrepancy may be due to our azimuthally averaging our data, while \cite{Tafalla416} did not.

\item L1512: The best match for this core is twice the canonical abundance of  ${\rm C}^{18}$O with significant depletion. This is consistent with work by \cite{Leeetal}. They give a greater ${r}_{d}$   than this work (15,000~AU instead of this work's 10,000~AU) but a lower ${f}_{d}$ (25 versus this work's 256). As in the case of L1498, Fig \ref{azavtwicecan} shows both findings to have similar $\chi^2_r$'s, and could be explained by our azimuthally averaging our data, while \cite{Leeetal} did not.

\item L1489: This core is well-modeled with canonical abundance of  ${\rm C}^{18}$O with significant depletion.

\item L1517B: Our results for L1517B show significant depletion, as does work by \cite{2004Ap&SS.292..347T}. Their work, however, found a best fit (to their observations) with ${f}_{d}$ $>$ 1000 and ${r}_{d}$ = 11,620~AU, which is more depletion than our work (see Table 2), perhaps because they use a T=10~K temperature profile, higher than ours.  

\item L1521E:  Our best fit for this core is twice our canonical abundance with ${r}_{d}$=2,500~AU and ${f}_{d}$=64. This represents little or no depletion, consistent with work by \cite{TafallaL1521E}, and suggests an early evolutionary state, consistent with work done by \cite{Hirotaetal}.  
Our exact findings for ${r}_{d}$ and ${f}_{d}$ , are not identical to those in this work, however, as \cite{TafallaL1521E} finds L1521E best fit by no depletion at all. However, \cite{TafallaL1521E} use a different density profile, and a different canonical abundance of $[{\rm C}^{18}$O]/[${\rm H}_{2}]$ (ours is $1.6\times10^{-7}$, versus $0.7\times10^{-7}$), and assumes a constant temperature T=10~K at all radii (Figure \ref {Tafallaprofilecompare}). Additionally, \cite{TafallaL1521E} observational values for the ${\rm C}^{18}$O 2-1 integrated intensity are 30\% higher than ours (Figure \ref{Tafallaprofilecompare}). Discussions with Tafalla have indicated that some of this discrepancy may be the result of different beam sizes, as the FWHM of his observations is 12$^{\prime\prime}$ versus our 34$^{\prime\prime}$, and systematic calibration errors (the IRAM observations were double sideband, while the HHT observations were single sideband).  

\item L1495A-S: Very high $\chi^2_r$, not well matched by any ${r}_{d}$ and ${f}_{d}$ with canonical abundance. Similar to TMC-2 in that it is best matched by a profile with no depletion if one assumes a canonical abundance. The lowest $\chi^2_r$ value, however, comes from doubling the abundance and assuming significant depletion.

\item L1544: Our results show this core to be best fit by using twice the canonical abundance but still significantly depleted, which is consistent with the conclusion of Lee et al (2004). They find an ${f}_{d}$ of 25 at an ${r}_{d}$ of 9,000~AU. This is more depletion than our findings, ${f}_{d}$ of 16 at an ${r}_{d}$ of 7,500~AU, however they are using a higher canonical abundance, $4.8\times10^{-7}$. \cite{Casellietal} found this core to have an abundance drop by a factor of approximately 10 at a radius of 6500~AU. Our Fig \ref{azavtwicecan} shows this result to have a higher $\chi^2_r$ than our best fits, but still fairly consistent with our results. \cite{Aikawa2001} find a central hole of 4,000~AU, inside of which no CO is present. This is more depletion than our findings, but their work assumes a higher ${\rm H}_{2}$ number density than ours, $1.5\times10^6$ ${\rm cm}^{-3}$. Our best fit value still has a $\chi^2_r$ greater than one, as seen in Table 2. This is likely due to L1544 being not round, but inspection of Figure \ref{ivsrfits} shows that we were able to find a combination of ${f}_{d}$ and ${r}_{d}$ that, at least by eye, appears to be a good fit.  

\end{enumerate}

\section{Analysis} We find that all cores but one, L1521E, are best fit by models that include significant depletion. Our conclusion that all but one of these cores is significantly depleted holds even taking into account reasonable variances in outer radius, temperature, and canonical abundance. In no case did we find varying any of these parameters substantially affected the results.  This work is unique in that we have tested a more complete parameter space of ${r}_{d}$, ${f}_{d}$, and C$^{18}$O canonical abundance than previously published work. While outer radius does not impact our results, choices of temperature profile and canonical abundance can. Even fixing these parameters, we cannot, however, robustly distinguish between a factor of 10 depletion and a factor of 1000
in ${f}_{d}$, as the $\chi^2_r$ are very similar (e.g., . Figures \ref{ivsrfits}, \ref{azavcan} and \ref{azavtwicecan}). In contrast, the depletion radius, ${r}_{d}$, for some of our cores is more strongly constrained than ${f}_{d}$; however, ${r}_{d}$ values for many of these cores are separated by little more than a beam width, and four of the cores have the same value for ${r}_{d}$.  Work by \cite{Leeetal} also found ${f}_{d}$ hard to constrain.

At the distance of Taurus, information in the HHT maps contain information on scales larger than $\theta_{mb}/2 = 2100$ AU.
Greater spatial resolution would better constrain ${r}_{d}$.  Improved resolution may even help constrain ${f}_{d}$. For instance, future observations with the Large Millimeter Telescope (LMT) 50-m operating at 1.3mm would increase our resolution by a factor of 5. The difficulty in determining the absolute value of ${f}_{d}$ limits our ability calculate an accurate chemical depletion time scale. The rate of freeze-out, or the chemical depletion timescale of CO, is given in \cite{diffyq} and \cite{Redmanetal}. The version in \cite{Redmanetal} gives ${\dot n}_{CO}\propto{\it n}_{CO}$. The solution of this differential equation implies that the amount of gas-phase CO decreases exponentially with time for a constant grain temperature, ${\it n}_{CO}\propto{\rm e}^{-t}$. The inability to tell the difference between ${\it f}_{d}$=10 vs ${\it f}_{d}$=1000 depletion translates into a factor of 4 difference in the timescales. As pointed out in \cite{Redmanetal}, for a core with {${\rm H}_{2}$ density of $2.8\times10^5$ (near the median central density of
cores modeled by Shirley \& O'Malia, in prep.), 90\% of the CO is depleted at 43,000 years, or approximately half the central density free-fall time scale. But if the CO is actually depleted by 99\%, this brings it to the free-fall time scale, and 99.9\% depletion would be longer than the free-fall time scale. The difference between ${\it f}_{d}$=10 vs ${\it f}_{d}$=1000 could mean the difference between a core being much older than the free-fall time scale, hence dis-favoring gravo-turbulent fragmentation as the dominant core evolution scenario, or being younger than the free-fall time scale, and
consistent with gravo-turbulent collapse timescales.

Better constraints on the amount of depletion will be important for constraining ambipolar diffusion timescales, which depend on ionization fraction. \cite{ionization} found that freeze-out of ${\rm C}^{18}$O can be accompanied by depletion of metal ions, significantly lowering the ionization fraction. In the case of L1544, they find that ionization fraction can be ${10}^{-9}$, which would put the ambipolar diffusion timescale on the order of the free-fall timescale. They argue that better constraints on amount of CO depletion could place constraints on the ionization fraction, so these issues could be coupled. This timescale could be shorter than the chemical evolution time scale (depending on the amount of depletion, as discussed earlier), which would imply ambipolar diffusion is not the dominant process. 

In Figure \ref{allvsrd}, we compare the best $\chi^2_r$ model parameters to the physical parameters of the cores (central density, central temperature, as well as density and temperature at best fit ${r}_{d}$ values). The plotted points indicate the best fit ${f}_{d}$ and ${r}_{d}$ (minimum $\chi^2_r$), as given in Table 2.  Error bars show the range of ${f}_{d}$ and ${r}_{d}$ that correspond to $\chi^2_r$ $<$ min($\chi^2_r$)+1. Figure \ref{allvsrd} shows that it is much more difficult to constrain ${f}_{d}$ than ${r}_{d}$. We find no strong correlations for any of these parameters with ${r}_{d}$, and none for ${f}_{d}$, although large error bars on ${f}_{d}$ make teasing out trends difficult. One would expect the chemical evolution to track the dynamical evolution; as a core contracts, more CO would freeze out. Our work suggests something more complicated, as the cores with greater central density are no more likely to be significantly depleted than those with low density.  Warmer cores are also no less likely to be depleted.  For instance, L1521E has one of the lowest dust temperature profiles yet has the least depletion.

Our findings are consistent with the conclusion that these cores may be evolving at different dynamical rates.  There are several possible explanations for why cores in Taurus could be evolving
at different rates.  For example, that the magnetic field strengths are different across Taurus. Since the magnetic pressure $\propto {\rm B}^{2}$, small changes in magnetic field would lead to large changes in pressure support. This would lengthen the collapse timescale until the cores become
supercritical.  Unfortunately, the magnetic field strengths in
Taurus starless cores are poorly constrained and the observed variation from dust polarization
measurements is over an 
order of magnitude (e.g., $\approx 10 \mu$G in L1498 to $> 100 \mu$G in L1544 
\citep{Kirketal2006,Crutcheretal2004}.

Rates of evolution may also be affected by turbulence. For instance, 
the external pressure is an important factor in determining the
gravitational stability of the core \cite{Bonnor}.
While the starless cores themselves have very narrow linewidths that are not dominated by turbulence, the turbulence outside the cores may play an important role in the bounding pressure.  The transition from thermal to turbulent medium can be
quite sudden in dense cores \citep{Pinedaetal2010}.
The amount of turbulent pressure varies across Taurus \citep{GoldsmithLSS}.   
This external pressure is in addition to the weight of cloud material surrounding the
cores \citep{Ladaetal2008}.  Observations of the surrounding turbulent 
environment (e.g. $^{13}$CO maps, \cite{GoldsmithLSS}) compared with observations of the  
transition between the coherent core and more turbulent surrounding medium 
(e.g. NH$_3$ maps with the new 7-beam receiver at the Green Bank Telescope) are needed to  elucidate the
importance of the bounding pressure and the role of turbulence on core evolution. Further information on the effects of magnetic fields and turbulence specific to Taurus can be found in \cite{LeeKim} and \cite{GoldsmithLSS}.

It is also possible that freeze-out is not as simple as we may expect, as desorption processes may be important. \cite{desorption} investigated the effects of desorption mechanisms such as X-rays heating,  direct impact of cosmic rays, UV irradiation, and exothermic reactions on dust grain surfaces (e.g., formation of ${\rm H}_{2}$). They dismiss X-ray effects, as there are not many X-rays in these types of molecular clouds, but find that the other three mechanisms can potentially be significant. Calculating the magnitude of these effects, however, depends on poorly constrained parameters, such as the composition of ice on the mantle and the size and shape of dust grains 

Worsening the problem is that \cite{desorption} are forced to use observations of CO freeze-out to constrain these parameters. As a result, if there is a substantial uncertainty between the amount of depletion as found in our work, then these parameters are even more poorly constrained. They conclude that freeze-out rates are still high (98\% for a quiescent core of density ${10}^{5}$) even with these effects; however, they admit a large amount of uncertainty. \cite{Aikawa2003} have also done work on surface grain interactions, which could also help in understanding desorption processes.  More work is needed in this area.

Velocity structures and dynamical evolution may be tied to chemical evolution in more indirect ways, as pointed out in \cite{JoanLee}. They compare chemical evolution in static and collapsing density profiles, and find that in prestellar cores, both give similar results. However, in a dynamic case, the material being pulled into the center of the core may be coming from an outer radius that has less depletion. 
A better understanding of these cores' dynamics and environments would be very useful. Work by \cite{Aikawa2005}, \cite{Lee2004}, and \cite{Leethesis} has made much progress on this topic. However, as pointed out by M. Tafalla (private communication, May 9, 2010), the linewidths observed in many of these cores are inconsistent with arguments that cores such as L1521E are simply collapsing too quickly to have time for chemical freeze-out. 

\section{Conclusions} We find that all of our cores but one, L1521E, are best fit by models showing significant depletion, consistent with previous work in the field suggesting that unevolved cores are rare. This would imply that the depletion process happens very quickly. The lack of unevolved cores is likely an observational bias, as chemically unevolved cores may also be lower density than those in our sample set.  Observations with \textit{Herschel} and SCUBA2 may find many more lower density cores that may also be chemically unevolved.

We find that assuming an isothermal temperature profile can reproduce the observations, although one needs further information to know the appropriate temperature to use. This is likely due to the small temperature variations seen in dust continuum radiative transfer models.
T=10~K, however, rarely provides a good best match.  We find assuming 
$T_d = T_{gas}$ gives similar results to energetics calculations with $T_d \neq T_{gas}$. Our work also finds no correlation between the amount of depletion and central density or central temperature, nor do we find correlations between density or temperature at the radius of depletion. It is possible that these cores are evolving at
different rates that may be affected by variations in the magnetic field strengths,
turbulence, and bounding pressures across Taurus.

Constraining the amount of depletion in these cores can be complicated by degeneracies between temperature and canonical abundance and difficulty in placing strong constraints on ${f}_{d}$. Constraining ${f}_{d}$ more strongly is necessary to compare chemical evolution time scales to free-fall time scales, an important step in distinguishing between the relative importance of ambipolar diffusion and gravo-turbulent fragmentation in the evolution of starless cores.
Ultimately, observations with better angular resolution, as well as additional understanding of the desorption processes, dynamics, and surrounding environments of these cores, will help complete our understanding of depletion and chemical evolution in starless cores.


\begin{acknowledgments} 
\begin{center} ACKNOWLEDGEMENTS \end{center}

We thank Amelia Stutz, Mario Tafalla, Jeong-Eun Lee, George Rieke, Joan Najita,  Kaisa Young, Neal Evans and Craig Kulesa for useful discussions and comments. We are additionally grateful to Floris van der Tak \& Michael Hogerheijde for gracious assistance in implementing their code. We also thank Andy Marble and Megan Reiter for help with figures.

This material is based upon work supported by the National Aeronautics and Space Administration through the NASA Astrobiology Institute under Cooperative Agreement No. CAN-02-OSS-02  issued through the Office of Space Science.

\end{acknowledgments}


\bibliographystyle{aa}
\bibliography{aford}

 \par\pagebreak
\begin{table}
\caption{Characteristics of These Cores}
\centering
\begin{tabular}{c c c c c c c c }
\hline\hline
Core & RA (J2000) & Dec & ${\rm v}_{LSR}$ & Central density & Tinner & Touter & $\sigma _{I({\rm C}^{18}O\, 2-1)}$ \\ [0.5ex]
 & & & km/sec & ${\rm cm}^{3}$ & K & K & K*km/sec \\[0.5ex]
\hline
TMC-2  & 04:32:45.5 & +24:25:07.7 & +6.30 &  $2\times10^4$ & 10.7 & 10.9 & 0.180\\ 
L1498 & 04:10:53.0 & +25:10:08.6 & +7.78 & $2\times10^4$ & 10.2 & 10.3  & 0.194 \\
L1512  & 05:04:08.6 & +32:43:24.6 &  +7.10 & $8\times10^4$ & 8.2 &  8.6 & 0.157 \\
L1489 &  04:04:47.6 & +26:19:17.9 & +6.85 & $1\times10^5$ & 11.7 &  12.4 & 0.166\\
L1517B  & 04:55:18.3 & +30:37:49.8 & +5.78 & $2\times10^5$ & 8.7 & 9.4 & 0.200 \\ 
L1521E & 04:29:14.9 & +26:13:56.6 & +6.90  & $3\times10^5$& 7.6 &  8.5 & 0.187 \\
L1495A-S & 04:18:39.9 & +28:23:15.9 & +7.29  &  $4\times10^5$ & 9.7 & 12.8 & 0.187 \\
L1544 & 05:04:17.2 & +25:10:43.7 & +7.18 & $8\times10^5$ & 8.5 & 11.4 & 0.166 \\
\hline
\end{tabular}
\label{coords}
\end{table}

\begin{table}
\caption{Best Fit:  rd, fd, and Abundances (as a multiple of the canonical abundance)}
\centering
\begin{tabular}{c c c c c }
\hline\hline
Core & ${r}_{d}$ & ${f}_{d}$& Abundance & $\chi^2_r$ \\ [0.5ex]
& AU & & &  \\ [0.5ex]
\hline
TMC-2  & 5000 & 512 & 2 & 0.094 \\
L1498 & 7500 & 16 & 1 & 0.063 \\
L1512  & 10000 & 256 & 2 & 0.13 \\
L1489 & 7500 & 1000 & 1 & 0.48 \\
L1517B  & 10000 & 32 & 2 & 0.087 \\
L1521E & 2500  & 64 & 2 &  0.18 \\
L1495A-S  & 7500 & 4 & 2 & 0.42 \\
L1544   & 7500 & 16 & 2 & 1.5 \\
\hline
\end{tabular}
\label{rdfd}
\end{table}


\begin{figure}
\includegraphics[width=0.9\textwidth]{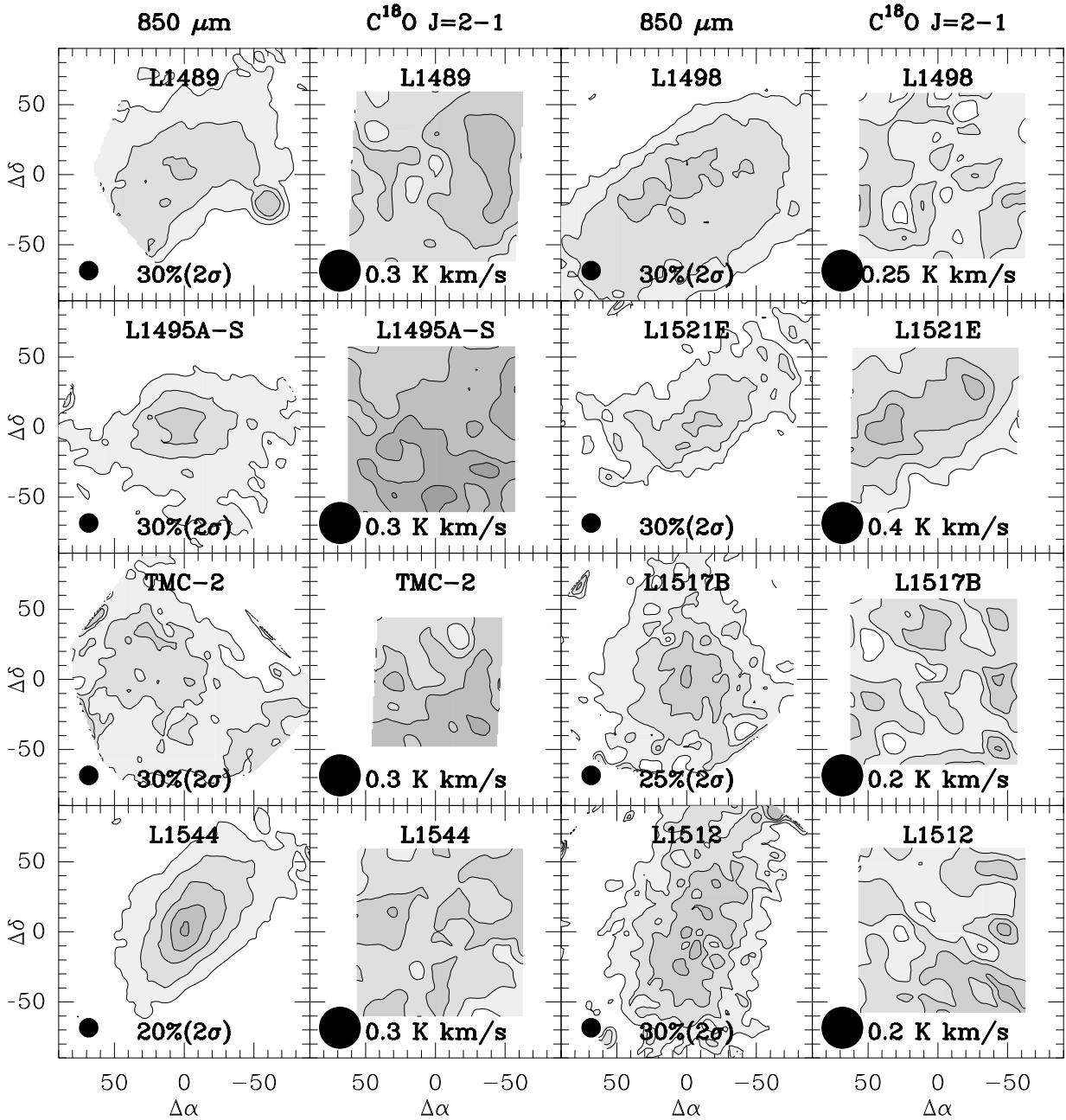}
\caption{On left, 850~\micron\ dust continuum emission, from SCUBA data, normalized (arbitrary units). On right, ${\rm C}^{18}$O  integrated intensity. Beam size indicated by solid circles. Contours are percentage of the peak for the 850~\micron continuum, and value indicated for the ${\rm C}^{18}$O integrated intensity. }
\label{dustemiss}
\end{figure}

\begin{figure}
\includegraphics{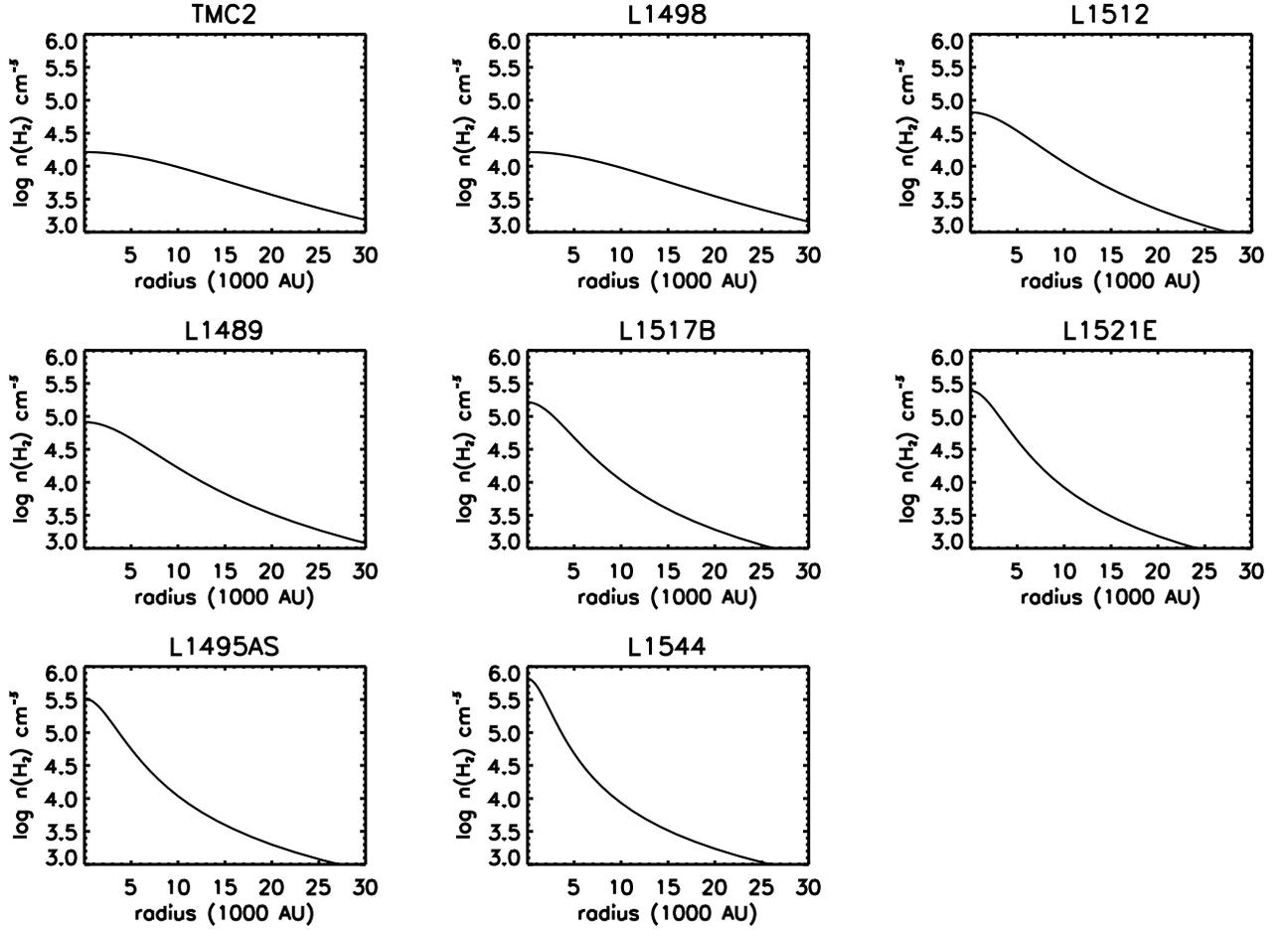}
\caption{${\rm H}_{2}$ number density vs. radius profiles used for each of the cores. }
\label{alldensity}
\end{figure}

\begin{figure}[h]
\includegraphics[width=0.80\textwidth]{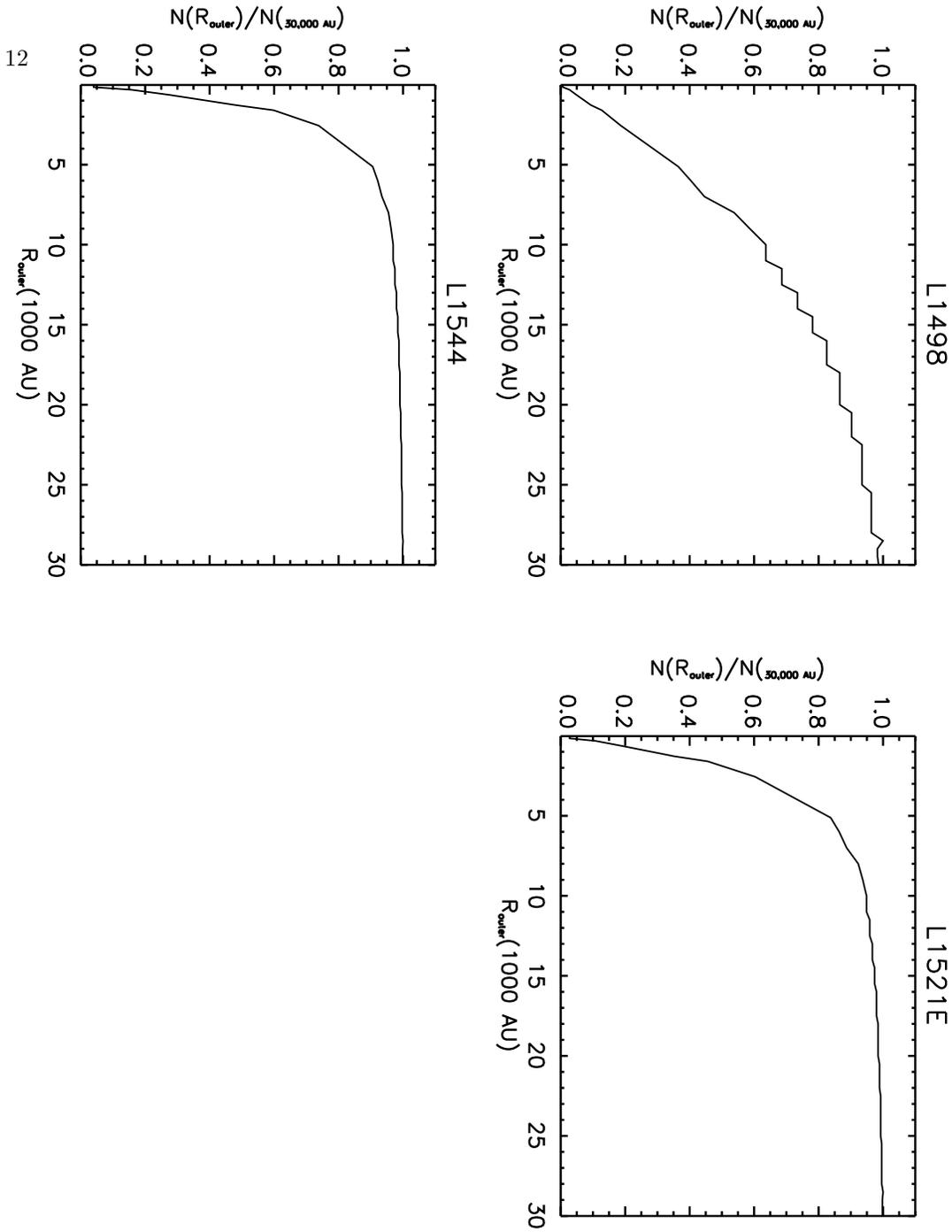}
\caption{We explore possible effects of choice of ${\rm R}_{outer}$.  We integrate the column density from r=0 (the center) to r=${\rm R}_{outer}$ (the outer radius, which we vary from zero to our model cutoff of 30,000~AU), then divide that value by the integrated column density from 0 to 30,000~AU, and plot that ratio as a function of ${\rm R}_{outer}$ . We do this for three model cores with densities of $2\times10^4$, $3 \times10^5$, and $8\times 10^5~{\rm cm}^{-3}$, representing the full span of our sample's central density. }
\label{nofr}
\end{figure}

\begin{figure}
\includegraphics[width=0.9\textwidth]{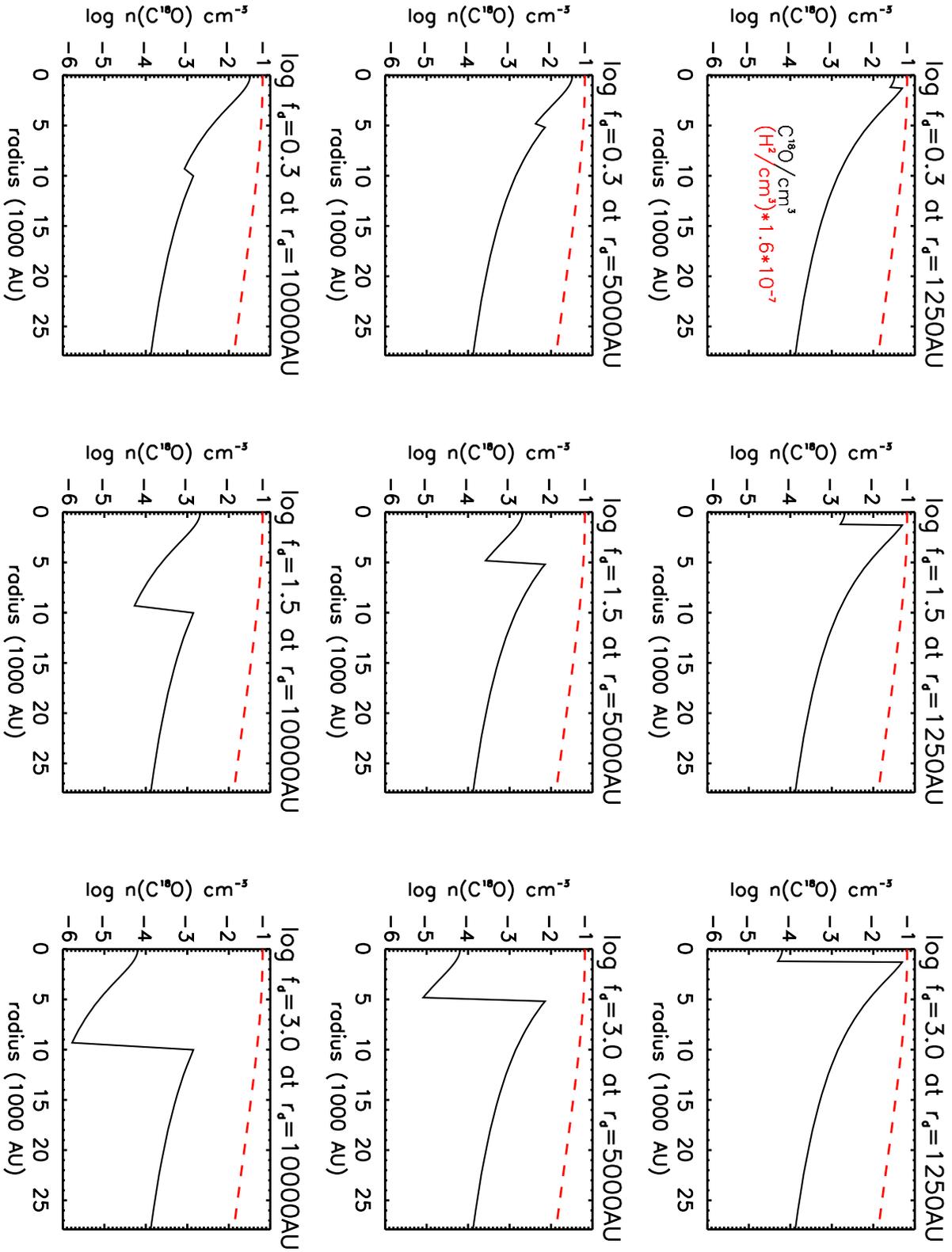}
\caption{Density of ${\rm C}^{18}$O vs. radius for representative combinations of ${r}_{d}$ (in AU) and ${f}_{d}$. Overplotted (red dashed line) is the density of ${\rm H}_{2}$, just shifted down by the canonical abundance, [${\rm C}^{18}$O]/[${\rm H}_{2}$]= ${1.6\times{10}^{-7}}$, so as to fit on the same plot. }
\label{covsr}
\end{figure}

\begin{figure}
\includegraphics[width=0.8\textwidth]{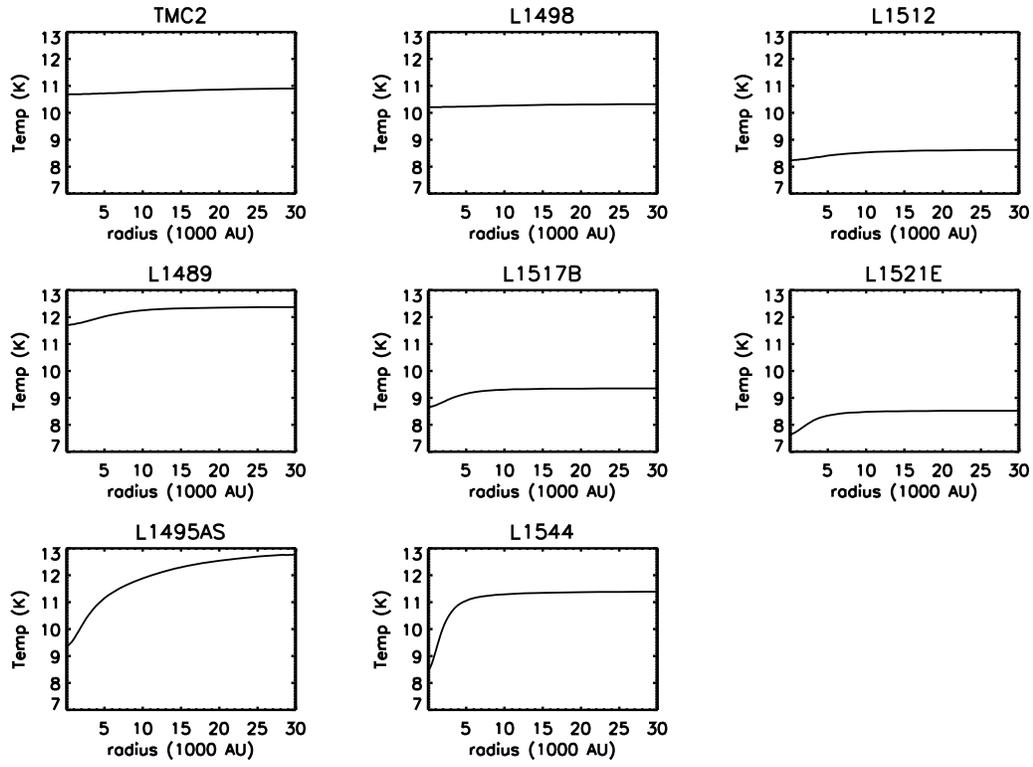}
\caption{Temperature vs. radius profiles from dust continuum radiative transfer models used for each of the cores.}
\label{alltemp}
\end{figure}

\begin{figure}
\includegraphics[width=0.8\textwidth]{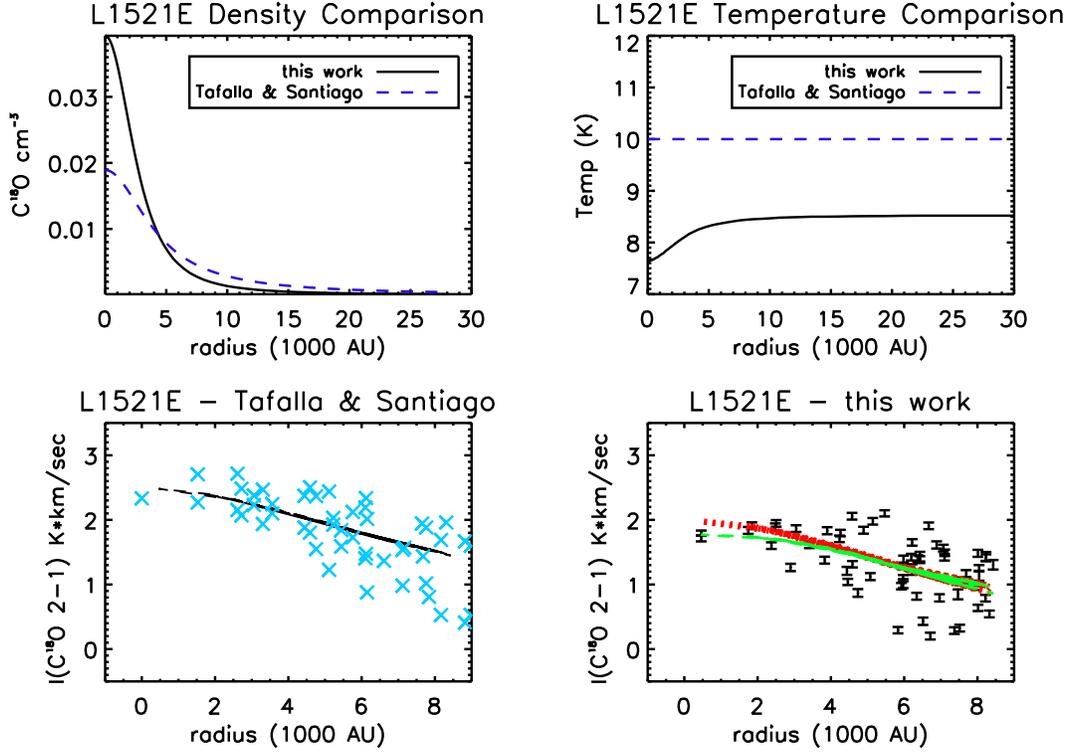}
\caption{For L1521E, comparison of temperature and density profiles used in \cite{TafallaL1521E} to those in this work. The bottom panels compare observed and model values of integrated intensity (K  km ${\rm sec}^{-1}$) of the ${\rm C}^{18}$O J=2-1 for both this work and theirs. In the bottom left panel, blue crosses are their observed points. The solid black line was made by using their temperature and density profiles, which we ran through RATRAN and our associated analysis. It appears to be identical to their best fit curve found in their Figure 3. In the bottom right panel, the black points are our observed points. The red (upper) line is for no depletion, with our temperature and density profiles (twice the canonical abundance), run through RATRAN. The green (lower) line is our overall best fit with ${r}_{d}$=2,500~AU and ${f}_{d}$=64. For further discussion, see \S~3.3.}
\label{Tafallaprofilecompare}
\end{figure}

\begin{figure}[h]
\includegraphics[width=0.80\textwidth]{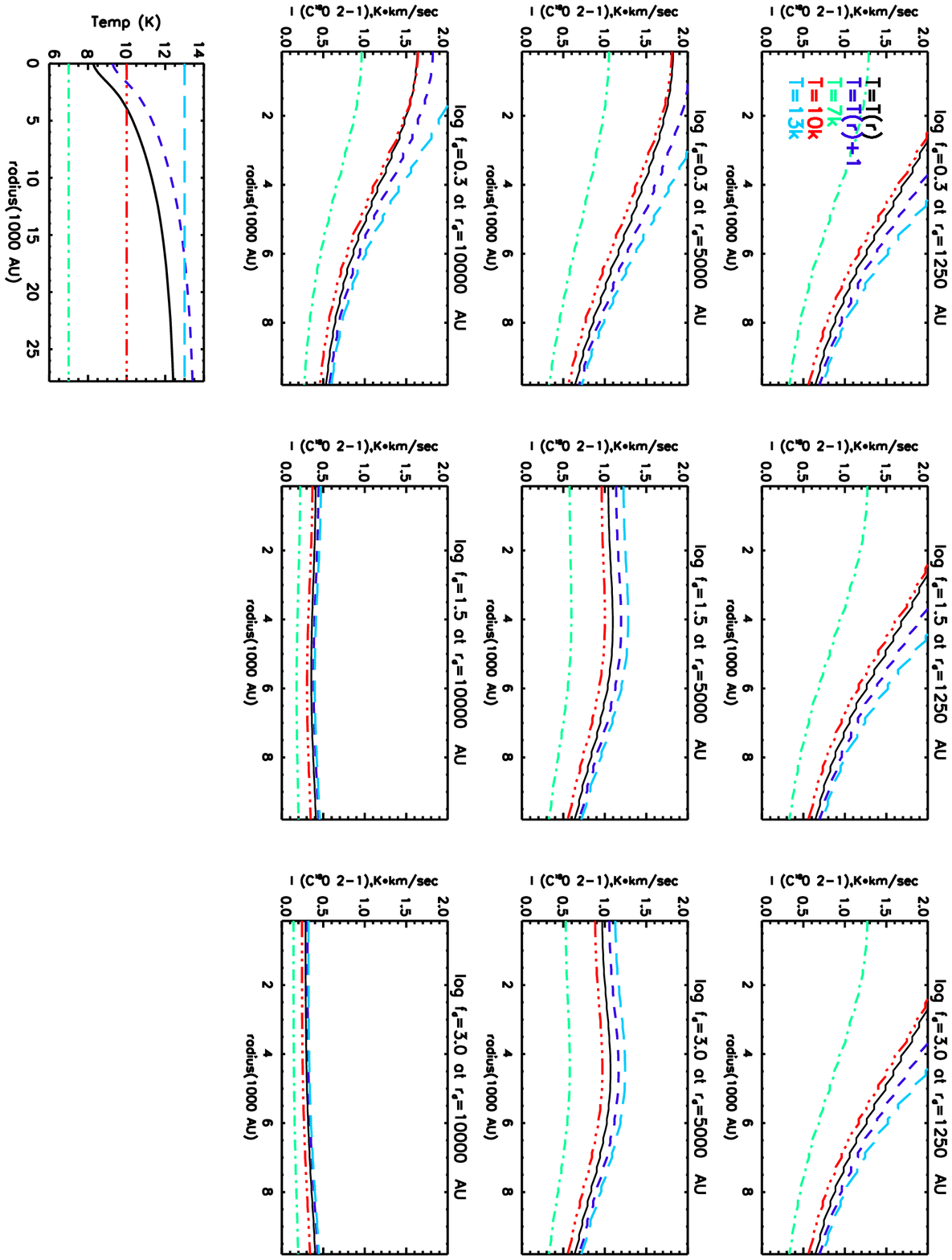}
\caption{Intensity vs. radius for various combinations of {\it ${r}_{d}$} and {\it ${f}_{d}$}. This is for a theoretical B-E sphere with an example density of $3\times10^5~{\rm cm}^{-3}$. The bottom panel shows the temperature profiles used in the nine panes above, with matching linestyles.}
\label{nc3e5plus1}
\end{figure}

\begin{figure}
\includegraphics[width=0.8\textwidth]{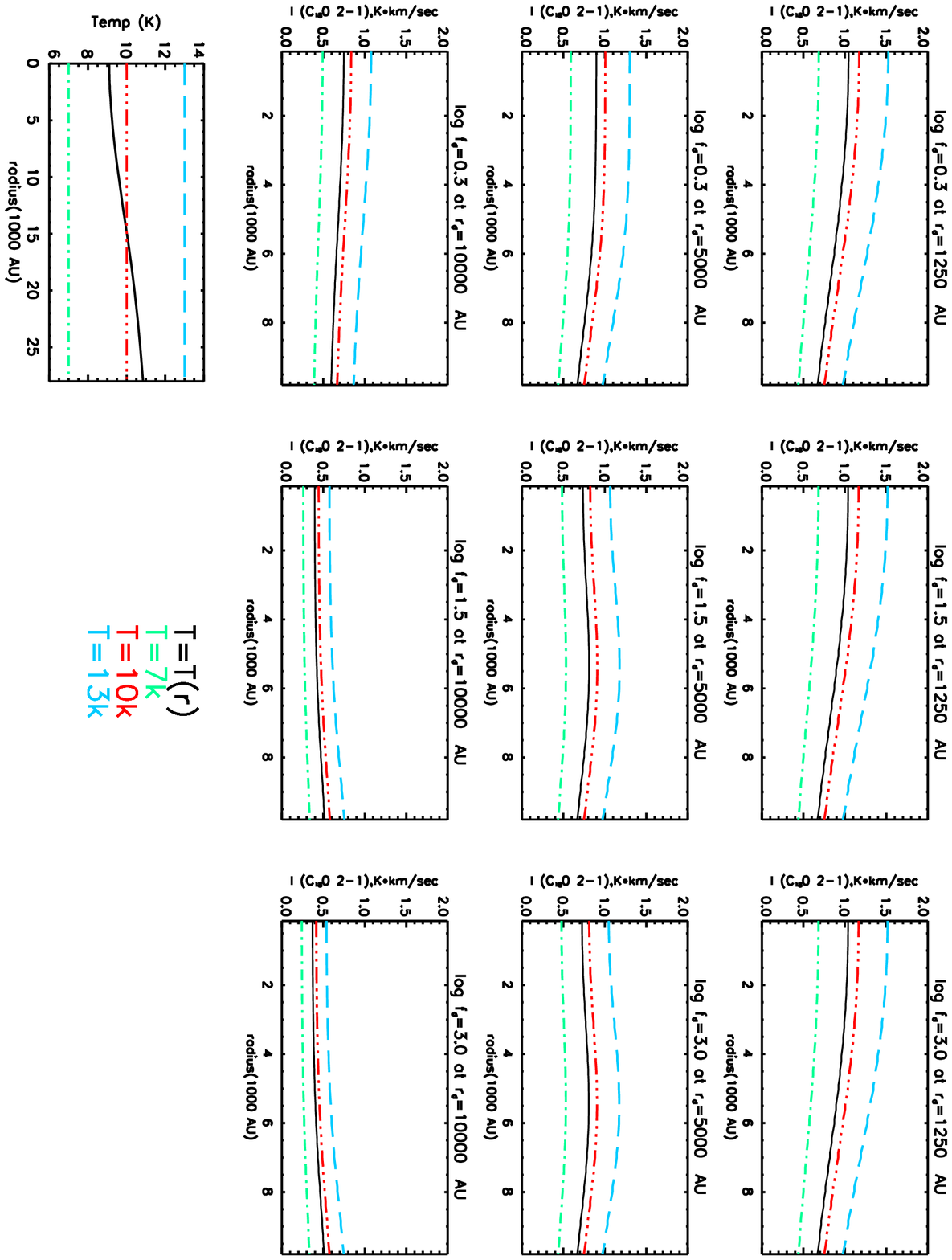}
\caption{Intensity vs. radius for various combinations of {\it ${r}_{d}$} and {\it ${f}_{d}$}. This is for a theoretical B-E sphere with a central density of $2.5\times10^4  ~{\rm cm}^{-3}$. The bottom panel shows the temperature profiles used in the nine panes above, with matching linestyles.}
\label{2pt5e4}
\end{figure}

\begin{figure}
\includegraphics[width=0.8\textwidth]{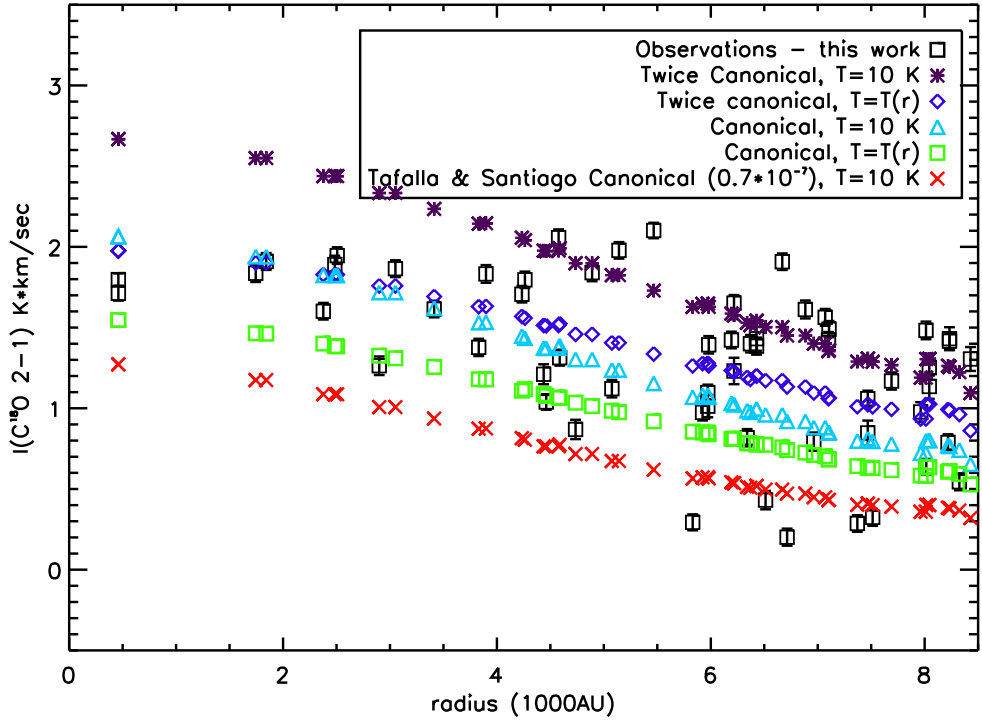}
\caption{Illustration of modeling degeneracies present in L1521E, as an example. Note that our data points (black) can be fit well with either our temperature profile, which varies with radius and double the canonical abundance of ${\rm C}^{18}$O, or with a constant temp profile of T=10~K and the regular abundance. Model points exactly match coordinates of observed points.}
\label{L1521Edegeneracies}
\end{figure}

\begin{figure}
\includegraphics[width=0.8\textwidth]{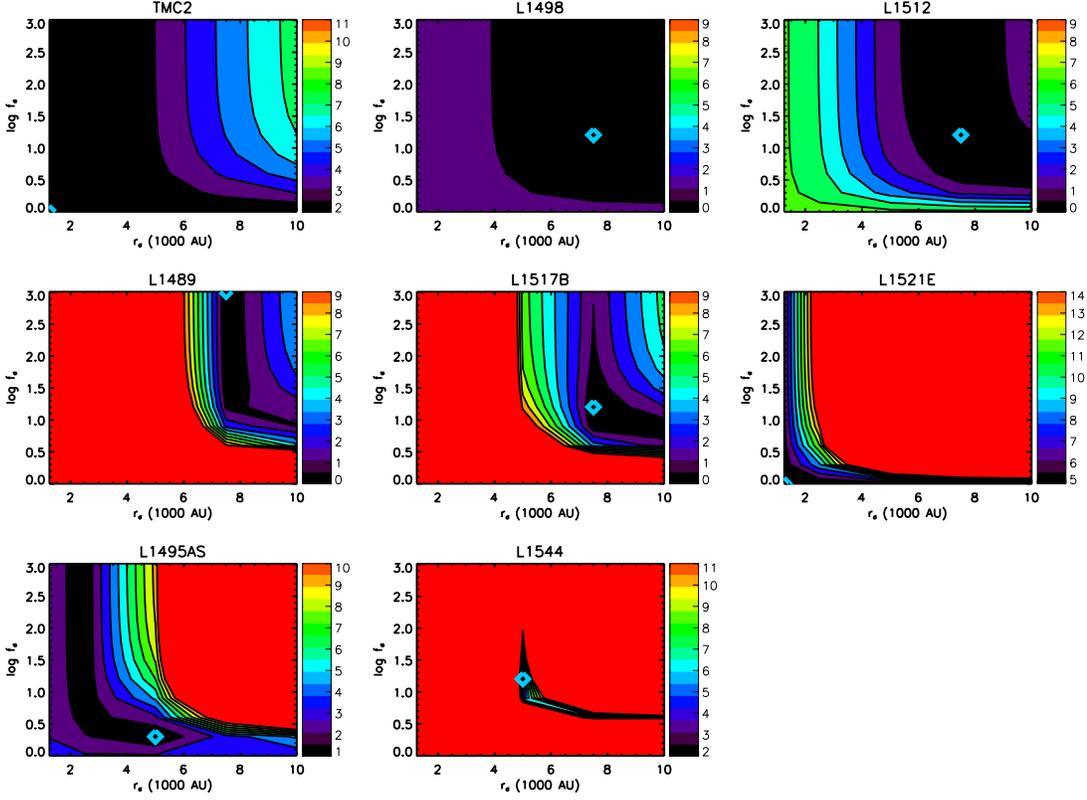}
\caption{$\chi^2_r$ maps of  each of our 8 sources, assuming a canonical abundance, $1.6\times10^{-7}$. Observed and model points have been azimuthally averaged in 10" annuli. The black regions represent  $\chi^2_r$  less than minimum($\chi^2_r$)+1. For TMC2, L1521E, L1495AS, and L1544, no combination of  ${r}_{d}$ and ${f}_{d}$ gives a $\chi^2_r$ less than 1. Note that the a wide range of ${f}_{d}$ gives very similar $\chi^2_r$ values, showing this value is poorly constrained. Blue diamonds represent the location of the lowest $\chi^2_r$, note they are in bottom left corner for TMC2 and L1521E.}
\label{azavcan}
\end{figure}

\begin{figure}
\includegraphics[width=0.8\textwidth]{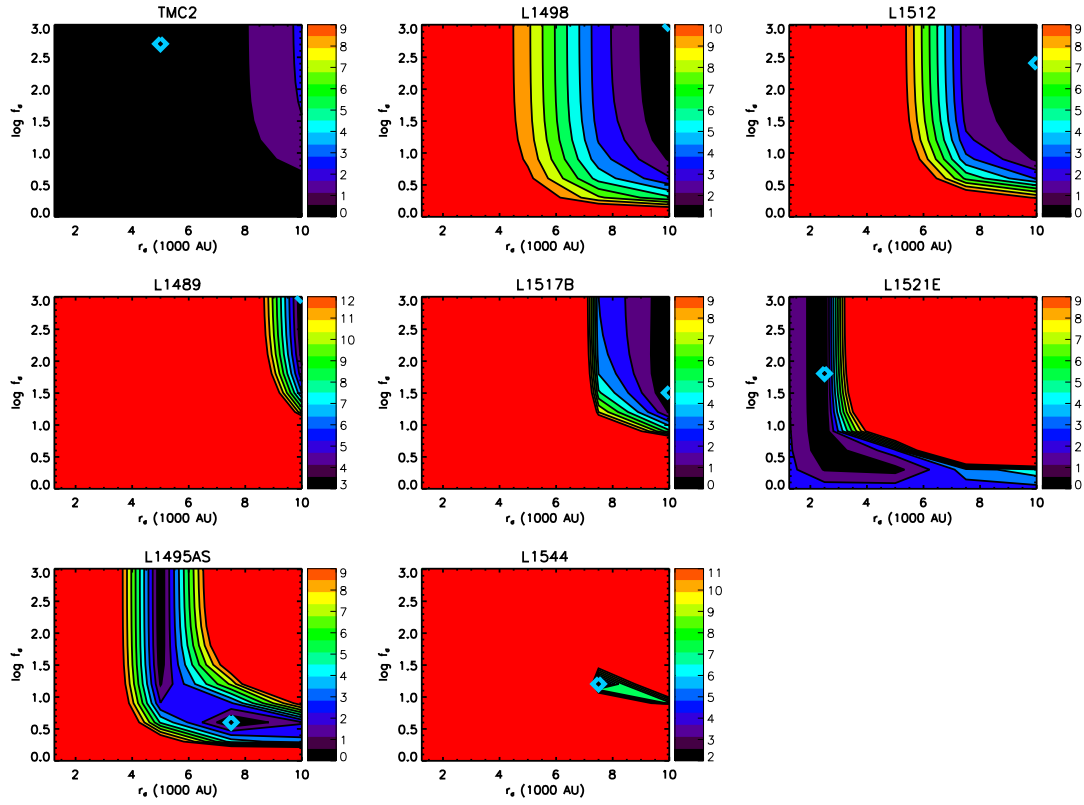}
\caption{$\chi^2_r$ maps of  each of our 8 sources, assuming twice the canonical abundance, or $3.2\times10^{-7}$.  Observed and model points have been azimuthally averaged in 10" annuli. TMC2, L1521E, and L1495AS, now have $\chi^2_r$ less than 1. ${f}_{d}$ is still poorly constrained. Blue diamonds represent the location of the lowest $\chi^2_r$, note they are in the upper right corner for L1498 and L1489.}
\label{azavtwicecan}
\end{figure}

\begin{figure}
\includegraphics[width=0.8\textwidth]{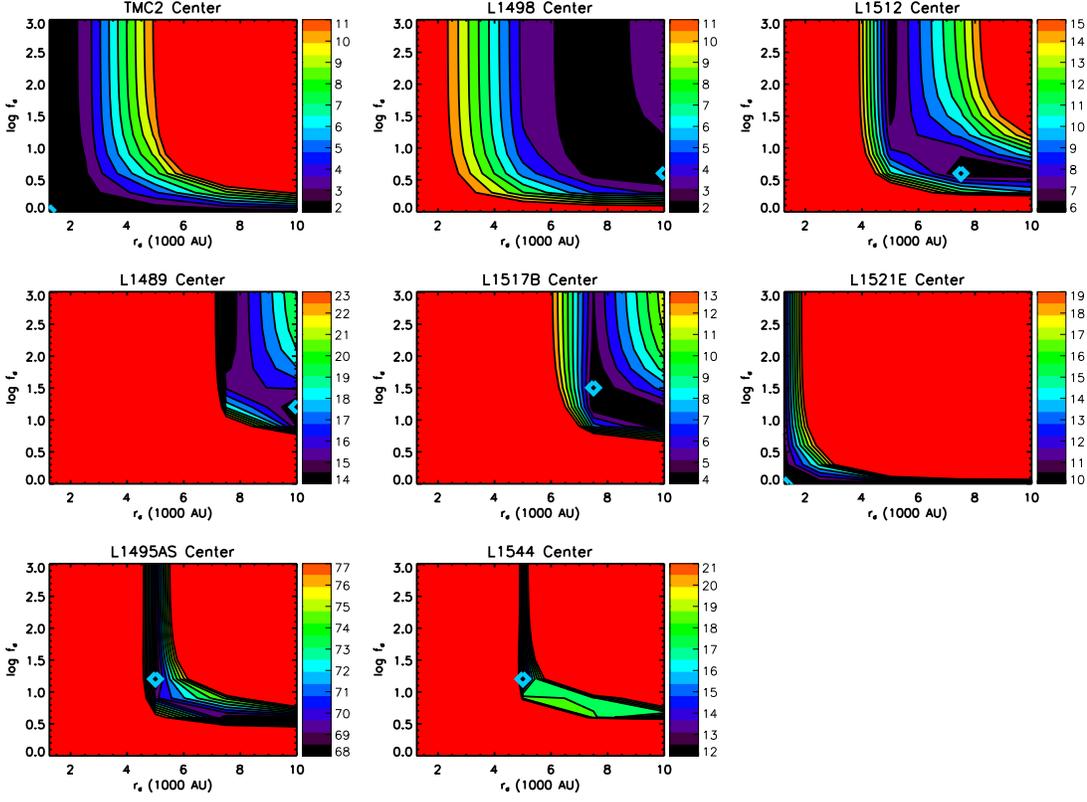}
\caption{$\chi^2_r$ of the central regions, showing the models that best match the central spectrum, assuming canonical abundance.  The contours (starting with the minimum $\chi^2_r$ value and increasing by 1) are similar in shape to Figure \ref{azavcan}, but with higher $\chi^2_r$ , as there is more noise in the line. L1495AS is not well fit due to the flat line shape, see Figure \ref{centerchisq}. Blue diamonds represent the location of the lowest $\chi^2_r$, note they are in bottom left corner for TMC2 and L1521E. Note this figure is not azimuthally averaged since it is focused only on the central region of each core.}
\label{centerchisqcontour}
\end{figure}

\begin{figure}
\includegraphics[width=0.8\textwidth]{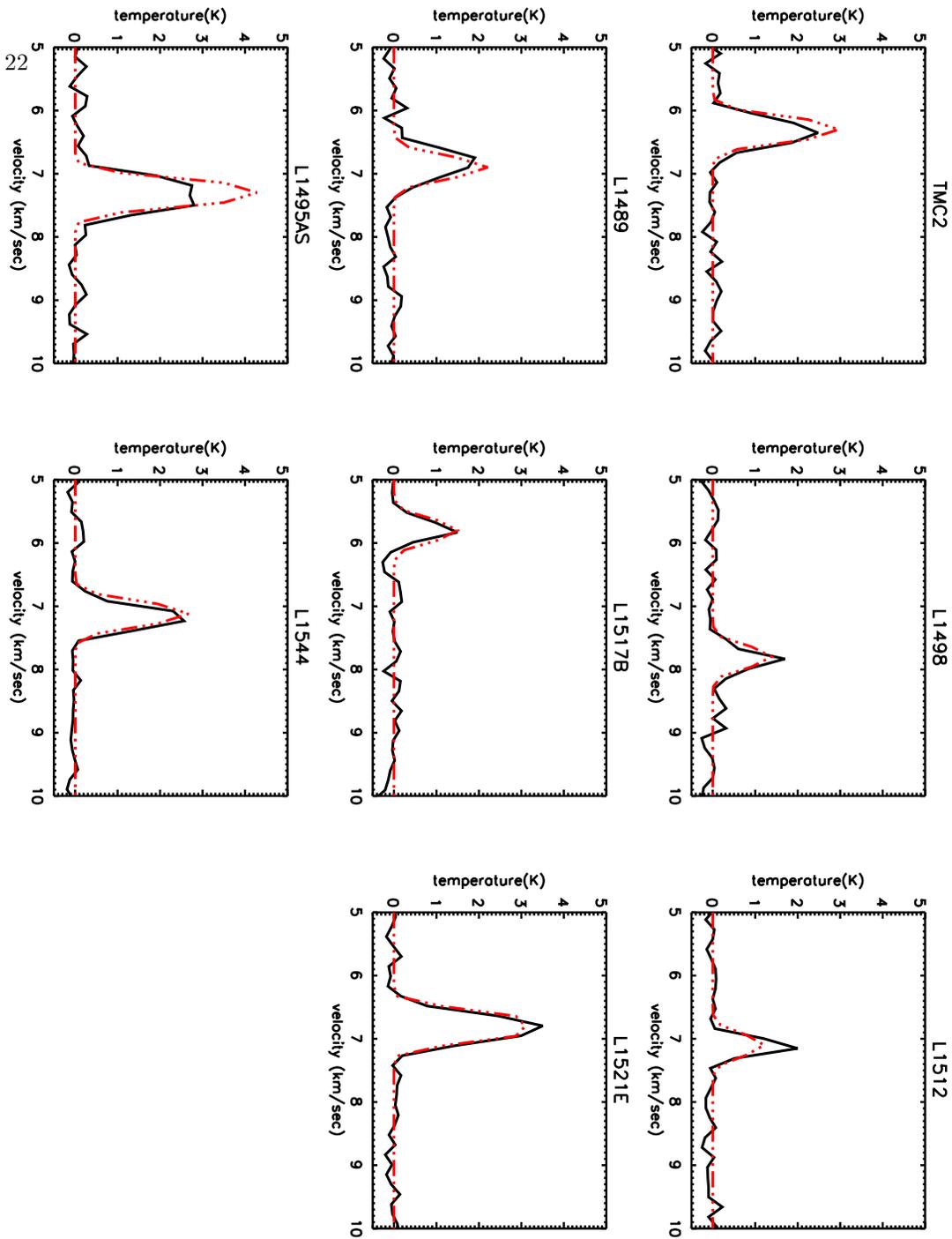}
\caption{Central spectra of each core, for both the observed values (black solid) and the model value (red dotted) with the lowest $\chi^2_r$ for  the whole region. See Table 2.}
\label{centerchisq}
\end{figure}

\begin{figure}
\includegraphics[width=0.9\textwidth]{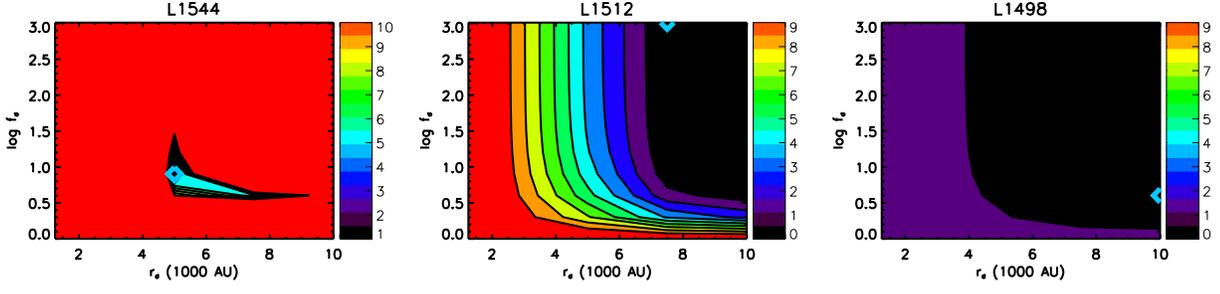}
\caption{$\chi^2_r$ maps of  cores for which work by \cite{Kaisa} found that gas and dust temperatures are not the same. Data and model points have been azimuthally averaged in 10" bins;  contour levels start with the minimum $\chi^2_r$ value. Blue diamonds represent the location of the lowest $\chi^2_r$.}
\label{Kaisa}
\end{figure}

\par\pagebreak

\begin{figure}
\includegraphics[width=0.8\textwidth]{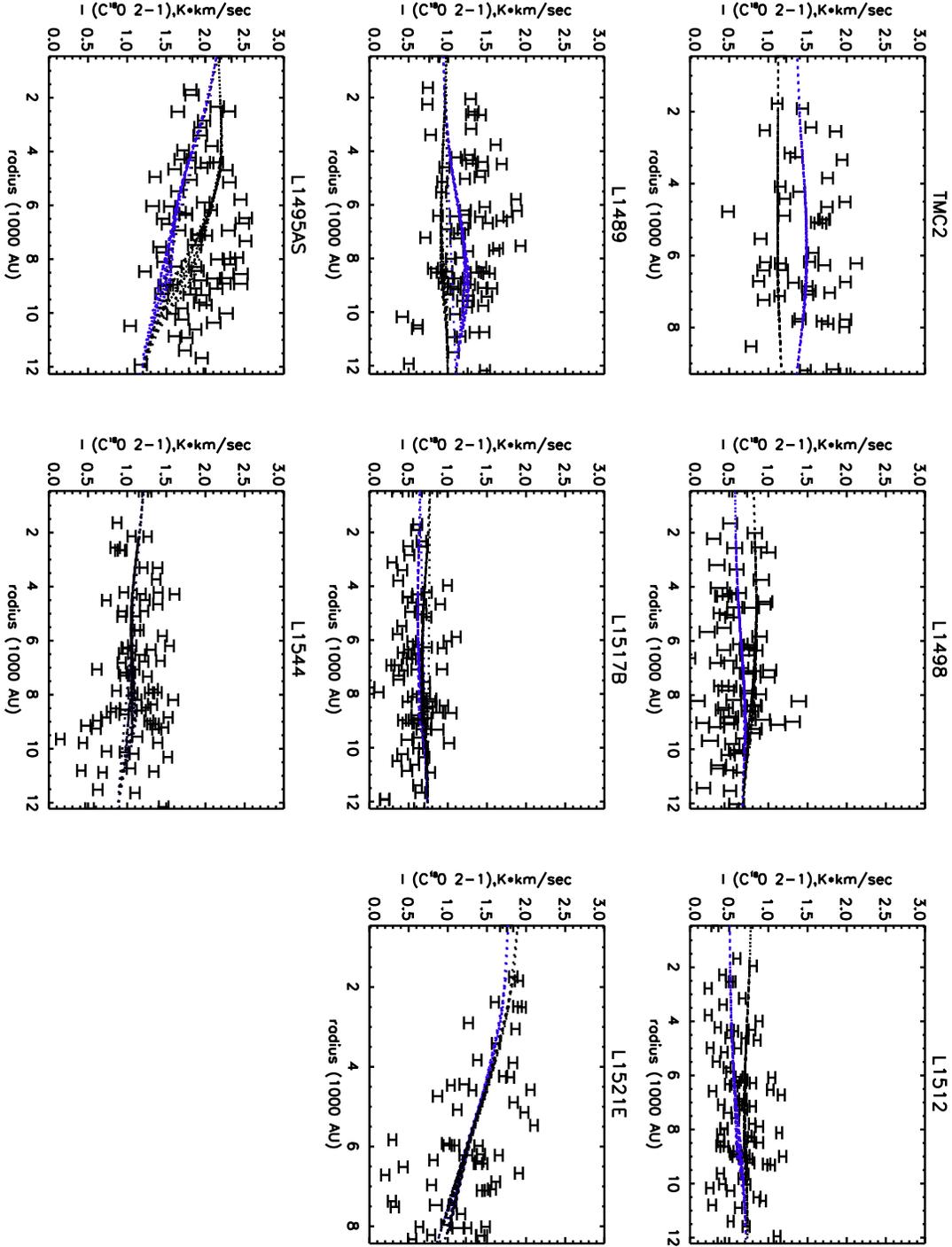}
\caption{Observational data (in black), as well as the combination of ${r}_{d}$ and ${f}_{d}$ that minimized the reduced $\chi^2_r$ (in blue). See Table 2. Errors in observed points are as defined in Equation 1. TMC2, L1521E, L1495A-S, L1517B, and L1544 are fitted with an abundance twice canonical. Plotted in black are the combination of  ${r}_{d}$ and ${f}_{d}$ that give the highest $\chi^2_r$, less than minimum($\chi^2_r$ )+1.  Black and blue lines are similar in most cases, illustrating the difficulty in constraining these quantities. See contour plots in Figs \ref{azavcan} and \ref{azavtwicecan}. This shows the difficulty in constraining ${f}_{d}$. For L1544, only one ${r}_{d}$ and ${f}_{d}$ combination gives $\chi^2_r$ $<$ min($\chi^2_r$)+1.}
\label{ivsrfits}
\end{figure}

\clearpage

\begin{figure}
\includegraphics[width=0.8\textwidth]{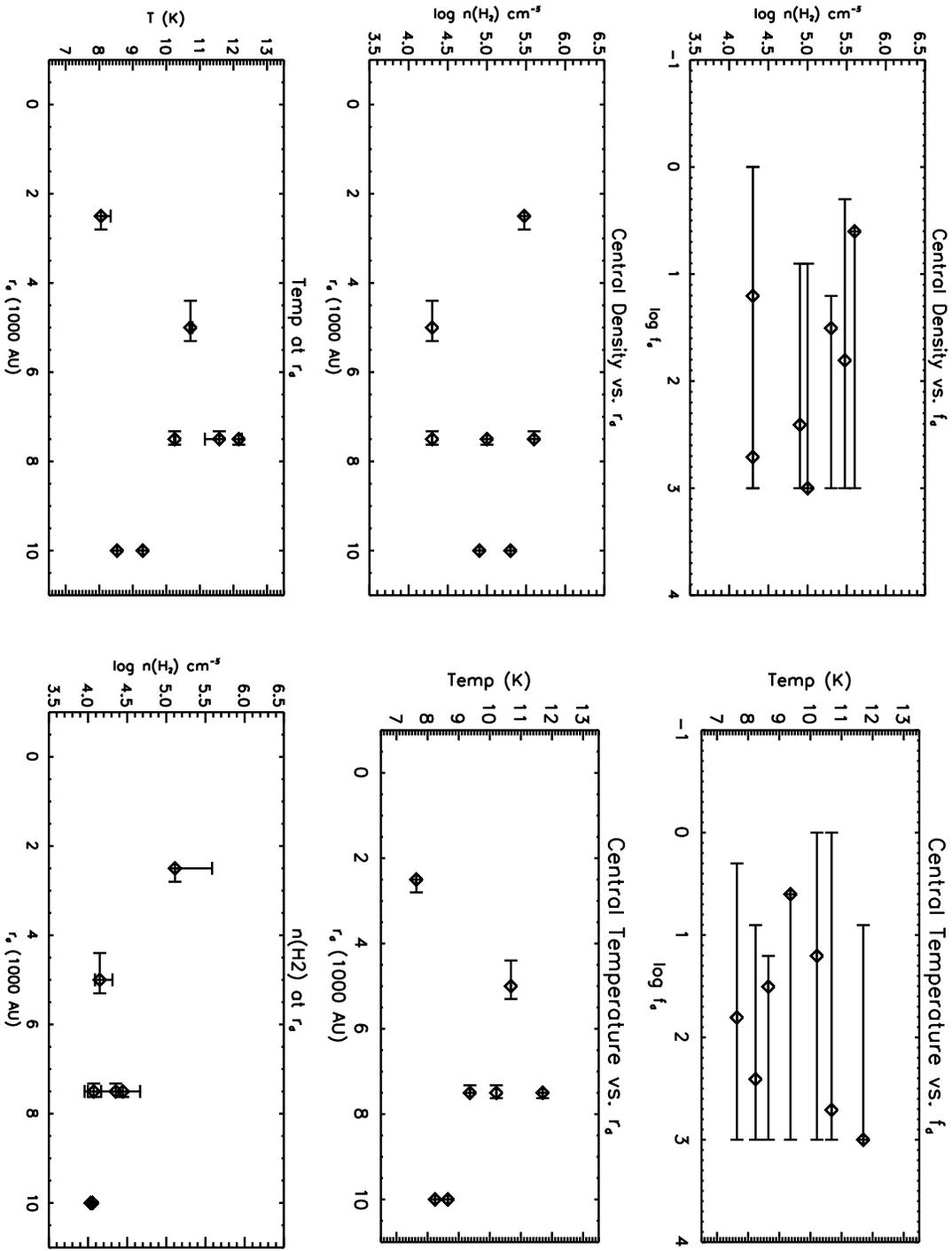}
\caption{We plot best fit ${f}_{d}$ and ${r}_{d}$ versus density and temperature at the centers of the cores. Error bars show the ranges of values that give $\chi^2_r$ $<$ minimum($\chi^2_r$)+1. We find no strong correlation of ${f}_{d}$ or ${r}_{d}$ with these parameters, although error bars for ${f}_{d}$ are large. This shows that ${f}_{d}$ is harder to constrain than ${r}_{d}$. The bottom two panes plot temperature and density for each core at the core's best fit ${r}_{d}$ values. These also show no strong correlation, as each core appears to be experiencing depletion at different temperatures and densities. L1544 does not appear in the figure, since no combination of ${r}_{d}$ and ${f}_{d}$ gave $\chi^2_r$ $<$1 for that core.} 
\label{allvsrd}
\end{figure}

\clearpage

\end{document}